\documentclass[pdflatex,sn-mathphys-num]{sn-jnl}

\usepackage{graphicx}%
\usepackage{multirow}%
\usepackage{amsmath,amssymb,amsfonts}%
\usepackage{amsthm}%
\usepackage{mathrsfs}%
\usepackage[title]{appendix}%
\usepackage{xcolor}%
\usepackage{textcomp}%
\usepackage{manyfoot}%
\usepackage{booktabs}%
\usepackage{algorithm}%
\usepackage{algorithmicx}%
\usepackage{algpseudocode}%
\usepackage{listings}%

\theoremstyle{thmstyleone}%
\theoremstyle{thmstyletwo}%

\theoremstyle{thmstylethree}%

\begin{document}

\title[Article Title]{AutoProsody: A Prosodic Feature Extraction Tool for Indian Languages}


\author*[1]{\fnm{Preethi} \sur{Thinakaran}}\email{preethit@snuchennai.edu.in}

\author[1]{\fnm{Malarvizhi} \sur{Muthuramalingam}}\email{malarvizhim@snuchennai.edu.in}
\equalcont{These authors contributed equally to this work.}

\author[1]{\fnm{Sooriya} \sur{S}}\email{sooriyass@snuchennai.edu.in}
\equalcont{These authors contributed equally to this work.}
\author[1]{\fnm{Anushiya} \sur{Rachel Gladston}}\email{anushiyarachelg@snuchennai.edu.in}

\author[2]{\fnm{Vijayalakshmi} \sur{P}}\email{vijayalakshmip@ssn.edu.in}
\author[1,3]{\fnm{Hema} \sur{ A Murthy}}\email{hema@cse.iitm.ac.in}

\author[1]{\fnm{Nagarajan} \sur{T}}\email{nagarajant@snuchennai.edu.in}

\affil*[1]{\orgdiv{Department of CSE}, \orgname{Shiv Nadar University Chennai}, \orgaddress{\street{Rajiv Gandhi Salai (OMR), Kalavakkam}, \city{Chennai}, \postcode{603 110}, \state{Tamilnadu}, \country{India}}}

\affil[2]{\orgdiv{Department of ECE}, \orgname{Sri Sivasubramaniya Nadar College of Engineering}, \orgaddress{\street{Rajiv Gandhi Salai (OMR), Kalavakkam}, \city{City}, \postcode{603 110}, \state{Tamilnadu}, \country{India}}}

\affil[3]{\orgdiv{Department of CSE}, \orgname{Indian Institute of Technology, Madras}, \orgaddress{\street{Adayar}, \city{Chennai}, \postcode{600036}, \state{Tamilnadu}, \country{India}}}


\abstract{The availability of prosodic information from speech signals is useful in a wide range of applications. However, deriving this information from speech signals can be a laborious task involving manual intervention.  Therefore, the current work focuses on developing a tool that can provide prosodic annotations corresponding to a given speech signal, particularly for Indian languages. The proposed AutoProsody tool provides time-aligned phoneme, syllable, and word transcriptions, syllable-level pitch contour annotations, break indices, and syllable-level relative intensity indices. The tool focuses more on syllable-level annotations since Indian languages are syllable-timed \cite{jeena2016syllables}. Indians, regardless of the language they speak, may exhibit influences from other languages. As a result, other languages spoken in India may also exhibit syllable-timed characteristics. The accuracy of the annotations derived from the tool is analyzed by comparing them against manual annotations and the tool is observed to perform well. Although the current work focuses on five languages, namely Tamil, Hindi, and Indian English, Malayalam and Gujarati the tool can be extended to other Indian languages and possibly other languages in syllable time as well.}




\maketitle

\section{Introduction}
\label{sec:intro}
Prosody is a broad term encompassing multiple features such as tones, intonation patterns, pitch accents, and rhythm. In this work, we use the term prosody specifically to refer to measurable acoustic-prosodic correlates: Pitch (F0 contour), Intensity (energy), Break Indices (pause-duration cues). While recent advancements in artificial intelligence have resulted in high-performing speech-based systems, these still do not accommodate prosodic variations in spontaneous speech or discourse. For instance, current text-to-speech (TTS) synthesizers produce speech that is highly intelligible and natural. However, these require a huge amount of training data and are still devoid of emotion and adequate prosodic variations. Incorporating prosodic information while training a TTS system could possibly improve the naturalness of synthetic speech, without necessitating an increase in the amount of training data. For instance, without prosodic cues, the sentence ``I can’t believe it'' sounds flat and monotone; 
with prosody, it exhibits natural pitch movement and stress that convey genuine surprise. 

Prosodic features influence speech processing in distinct ways:
\begin{itemize}
    \item \textbf{Pitch (F0)} distinguishes interrogatives from declaratives and conveys speaker emotion.
    \item \textbf{Relative intensity (RI) } correlates with stress and prominence, also supporting emotion recognition.
    \item \textbf{Break Indices (BI)} classify inter-word silences into short, medium, and long pauses reflecting boundary strength.
\end{itemize}

\noindent\textbf{Example 1 — Pitch:}  
``You’re coming.'' vs.\ ``You’re coming?''  
Although the words are identical, a rising pitch contour in the second utterance signals a question.

\noindent\textbf{Example 2 — Intensity:}  
``I didn’t \textbf{say} he stole the money.''  
vs.  
``I didn’t say \textbf{he} stole the money.''  
Shifting intensity (stress) changes the emphasized word and alters the implied meaning.

\noindent\textbf{Example 3 — Break Indices:}  
``Old men $|$ and women were seated.'' (medium BI between men and women)  
vs.  
``Old men and women were seated.'' (short BI between men and women). The pause changes how the sentence is spoken and understood.

 Another application that can benefit from the availability of prosodic information is speech-to-speech (S2S) translation. An S2S system involves an automatic speech recognition (ASR) system that provides text corresponding to a given speech signal. This text is then translated to the desired language, which is then synthesized using a TTS system. Incorporation of prosodic information along with the ASR output could aid in producing speech in the target language that carries the same tones information as the speech in the source language. This is precisely where we have a problem.  Also when the order or sequence of word changes, incorporation of prosody in the target language is a big challenge. Further, each language has a unique prosodic structure and hence prosodic features could aid in language identification and in the incorporation of language specific prosody in TTS and S2S systems as well.

While prosody is immensely useful in a wide range of applications, obtaining accurate prosodic annotations is usually a laborious task requiring manual intervention. Further, while a standard representation of prosody exists for English, namely ToBI (Tones and Break Indices) \cite{Leben_1998,hirschbergTobi}, as well as for certain other languages, no such standard has been established for Indian languages. Therefore, the current work extends on the existing ToBI standard while incorporating additional information, namely time-aligned phonetic, syllabic, and word transcriptions and a relative intensity index, to develop a tool for prosodic annotation.

The ToBI system is a set of conventions for transcribing and annotating the prosody of speech, originally designed for English. Since the manual annotation of prosody is laborious, certain automatic labeling tools have been developed for English. One such tool is the Automatic ToBI (AuToBI) \cite{autobi}, which uses machine learning to predict ToBI labels from acoustic features of the speech signal. Another commonly used platform for annotation is praat \cite{parselmouth}. Although originally designed for manual annotation, it hosts several plugins and scripts supporting semi-automatic prosodic annotation \cite{praat}. Web MAUS \cite{webmaus} is a web-based service for automatic segmentation and labeling of speech, including prosodic features. Another popular tool is PyToBI \cite{pytobi}, which provides a Python interface for automatic labeling with ToBI. However, these are specifically designed for English.

The ToBI framework has now been extended to several other languages as well. For example, J-ToBI \cite{jtobi} has been designed to capture special pitch accent patterns and intonation structures in Japanese. Similarly, G-ToBI \cite{gtobi}, Sp-ToBI \cite{sptobi}, F-ToBI \cite{ftobi}, and C-ToBI \cite{ctobi} have been designed for German, Spanish, French, and Cantonese respectively. These adaptations indicate the relevance of ToBI across languages and the ease of tailoring it to different languages. 

Efforts to study prosody in Indian languages—characterized by their syllable-centric structures—have largely focused on analyzing \textit{intonation}, \textit{stress}, and \textit{rhythm} in languages such as Hindi, Bengali, Tamil, Kannada, and Telugu \cite{hindi_int, ind_intonation, bengali_pro}. However, a standardized framework for representing prosodic features across these languages is yet to be established. Recognizing the similarities in their prosodic structures, this work proposes a common prosodic annotation standard and tool. The tool automates the annotation of pitch contours labels, and break indices for Indian languages, while incorporating additional features, including:

\begin{itemize}
    \item Phoneme, syllable, and word boundary segmentation.
    \item Relative intensity index estimation.
\end{itemize}



We propose an automatic prosody tool specifically designed to extract prosodic features from Indian language speech. Unlike existing ToBI-based systems, this tool does not aim to assign categorical ToBI labels. Instead, it focuses on extracting continuous prosodic features—such as pitch contours labels, energy, duration measures, and syllable-level timing—which are more flexible and suitable for analyzing the diverse prosodic structures found in Indian languages. The tool is modular, language-agnostic, and scalable, enabling both exploratory prosodic analysis and downstream applications in speech synthesis, recognition, and corpus annotation. 

The paper is organized as follows: Section \ref{sec:prosody} provides a detailed description of the proposed tool, Section \ref{sec:perf} evaluates the accuracy of the annotations provided by the tool, Section \ref{sec:lid} presents the use of the tool to perform language identification, and Section \ref{sec:conc} concludes the paper.

\section{ The AutoProsody Tool}
\label{sec:prosody}

\begin{figure}[htbp]
    \centering
    \includegraphics[width=0.8\textwidth]{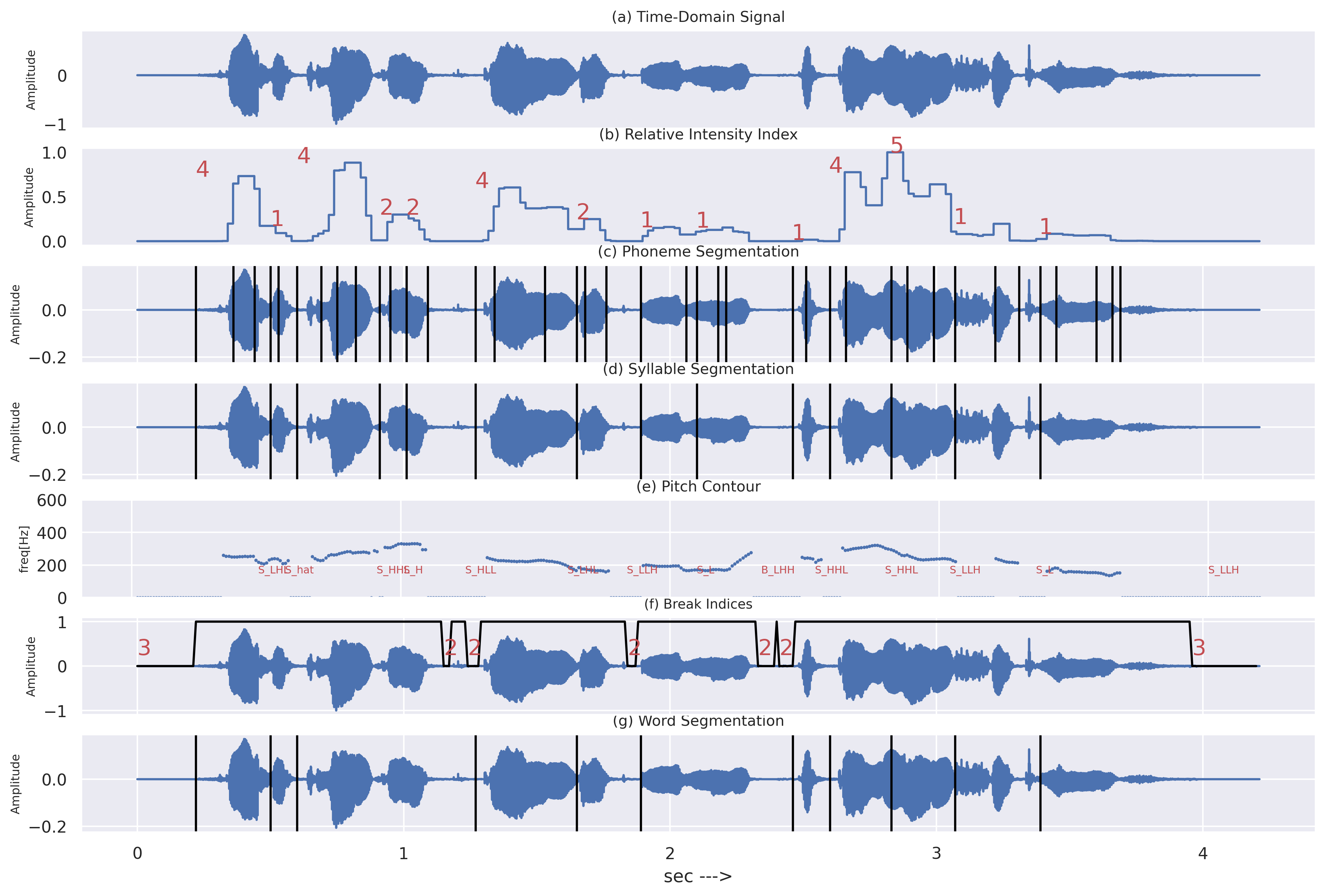}
    \caption{Annotations from the AutoProsody Tool for the English sentence, "For the twentieth time that evening the two men shook hands.''}
    \label{fig:SI-ToBI}
\end{figure}

The AutoProsody tool is designed to provide rich prosodic information from a speech signal. Given a speech signal, the tool converts it to a single channel and resamples it to a sampling frequency of 16 kHz. It then uses an ASR to obtain the corresponding orthographic transcription. This transcription, along with the speech signal, is used to derive phoneme, syllable and word-level boundaries. The tool also processes the speech signal to estimate the relative intensity index, pitch contour labels, and break indices. The extracted information is then plotted as shown in figure \ref {fig:SI-ToBI}.

In the AutoProsody tool, an automatic speech recognition (ASR) system is used when only the speech signal (WAV format) is available, without accompanying text. In this scenario, the ASR generates an orthographic transcription from the audio. The other parameters are extracted as mentioned above. However, it's important to note that the transcription produced by the ASR may sometimes be inaccurate, potentially leading to errors in the segmentation of the speech signal and in the extraction of other features.

Conversely, when both the speech signal (WAV) and text are available then the ASR system is unnecessary. In this case, language-independent models are directly applied to align the provided transcription with the speech signal, ensuring more accurate segmentation without the risk of errors introduced by the ASR.

Importantly, in both cases—whether the ASR system is used (when only audio is available) or not (when both audio and text are provided)—the AutoProsody tool consistently applies language-independent models.

Considering the number of languages in India, building systems for each language individually is a complex and time-consuming task. For instance, creating a Text-to-Speech (TTS) system for a new language requires a thorough understanding of its phonotactics (sound structure), letter-to-sound (LTS) rules, and other linguistic elements. This process involves considerable effort and often requires the expertise of linguists familiar with the language. As Prakash et al. (2014) highlighted in their work, "An approach to building language-independent text-to-speech synthesis for Indian languages" \cite{prakash2014approach},\cite{prakash2023tts}, building such systems from scratch for each language involves challenges such as training Hidden Markov Models, developing speech context modeling, and collecting language-specific data. These tasks are not only complex but also time-consuming.

By adopting language-independent models, the AutoProsody tool eliminates the need for such language-specific efforts. These models offer the flexibility of adapting to new languages with minimal modification, making the process more efficient. They do not require the extensive linguistic resources that traditional, language-dependent models need. Instead, they rely on generalizable algorithms that can be easily applied across multiple languages, greatly improving the scalability and applicability of the AutoProsody tool for the multilingual context of India.

This approach offers the advantage of easy adaptation to new languages compared to language-dependent models, thereby improving the accuracy of transcription processing and speech segmentation, regardless of the availability of text.

\begin{figure}[htbp]
    \centering
    \includegraphics[height=6cm, width=13cm]{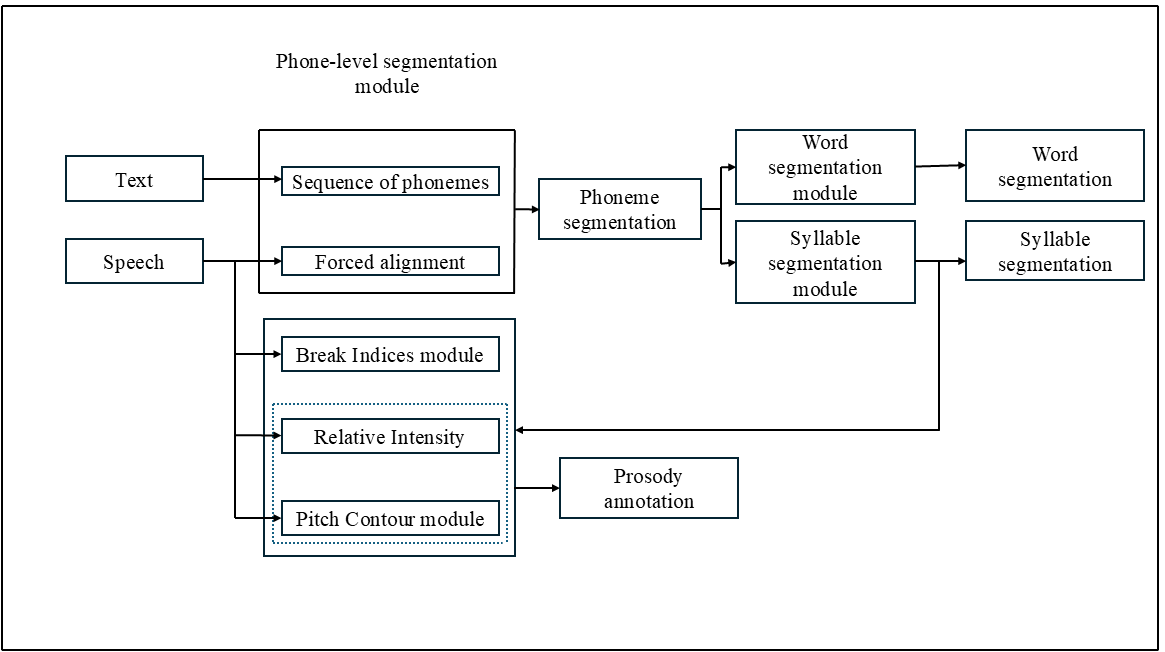}
    \caption{Block diagram of AutoProsody Tool}
    \label{fig:bd}
\end{figure}

The block diagram of AutoProsody tool figure \ref{fig:bd} outlines the overall process of text and speech processing, where input speech and text are segmented  and are processed via prosody modules.

\subsection{Automatic Speech Recognition}
\label{sec:asr}
To obtain the orthographic transcription corresponding to a given speech signal, the speech signal and the language (obtained from the user) are given to an ASR. In the current work, a pre-trained Data2Vec-aqc-based ASR model \cite{data2vec}, trained on 30,000 hours of speech data from 24 Indian languages, including Tamil, Hindi, and Indian English, Malayalam and Gujarati is employed.

\subsection{Segmentation}
\label{sec:segmentation}
Once the text corresponding to a given speech signal is obtained from the ASR, it is used to generate time-aligned phonetic, syllabic, and word transcriptions. To achieve this, context-independent hidden Markov models (HMMs) are trained for each phoneme across the five languages considered \footnote{Due to the requirement for less data and faster training, we favor HMM-based training over neural network methods.}. Phoneme segmentation is performed using HMM-based forced alignment. The input speech is first converted into acoustic features, and the corresponding phoneme sequence is obtained using a grapheme-to-phoneme (G2P) converter. Given the phoneme sequence and trained HMM acoustic models, forced alignment is applied to determine the start and end time of each phoneme in the speech signal. These HMMs are trained using three states, with one to five mixture components per state, depending on the number of available examples for each phoneme. The forced-Viterbi alignment procedure \cite{ijst, boothalingam2013development} is initially applied to identify phoneme boundaries, and this process is repeated over four iterations to obtain the final boundaries. This iterative process ensures the accurate segmentation of speech data, enabling robust phoneme, syllable, and word-level boundary derivation. Monophone HMMs for all phonemes are generated using the label files obtained. With these models, forced-Viterbi alignment is performed iteratively to segment the rest of the data, as described below:
\begin{enumerate}
    \item Using 5 minutes of speech data and the corresponding time-aligned phonetic transcriptions, context-independent phoneme models are trained (isolated-style training).
    \item Using these models and the phonetic transcriptions, the entire speech dataset is segmented using the forced-Viterbi alignment procedure.
    \item Using the newly derived time-aligned phonetic transcription (phone-level label files), new context-independent phoneme models are trained.

    \item Steps 2 and 3 are repeated for $N$ iterations ($N=5$ in this case).
    \item After $N$ iterations, the resultant HMMs are used to segment the entire speech data again. These boundaries are considered the final boundaries.
\end{enumerate}

This iterative process ensures the accurate segmentation of speech data, enabling robust phoneme, syllable, and word-level boundary derivation. Since the tool is designed to support all five languages, language-independent phoneme models are also trained, where acoustically similar phonemes across the languages share the same model. 

Syllabification involves combining phoneme boundaries to define syllable boundaries using a set of predefined rules. In this process, each vowel acts as the nucleus of a syllable, with consonants appearing at the beginning or end. The phoneme sequence is grouped into syllables in forms such as \( V \), \( C^*V \), \( VC^* \), and \( C^*VC^* \). This systematic approach ensures precise syllable segmentation, facilitating subsequent linguistic and prosodic analysis.

 While deriving the phoneme boundaries, it is ensured that there are silences inserted at the end of each word, though their duration might be close to zero at several places. These silences are used to combine the phoneme/syllable boundaries to derive word boundaries as discussed in 
Section \ref{sec:segmentation}.


\subsection{Computation of the Relative Intensity Index}
\label{sec:rii}
The tool then calculates the relative intensity index at the syllable level. The speech signal is divided into overlapping frames of 20 ms with a 10 ms hop length, and the Short-term energy (STE) is computed for each frame. The energy is subsequently normalized across the entire utterance. Based on the empirical analysis  for this normalized energy, a relative intensity index with difference of 0.2 is assigned to each syllable on a scale from 1 to 5 as follows:
\begin{equation}
	RII =
	\begin{cases} 
		1, & \text{if } E_N < 0.2 \\
		2, & \text{if } 0.1 \le E_N < 0.4 \\
		3, & \text{if } 0.4 \le E_N < 0.6 \\
		4, & \text{if } 0.7 \le E_N < 0.8 \\
		5, & \text{if } 0.9 \le E_N \le 1.0
	\end{cases}
\end{equation}
Fig. \ref{fig:SI-ToBI} (b) depicts $RII$, where the syllable, ``mehn'' has the highest RII of 5, while the syllables, ``dha'', ``iyv'', ``ningh'', ``shuck'', ``haendz'', etc. have the lowest RII of 1.

\subsection{Estimation of Break Indices}
\label{sec:bi}
In order to determine the break indices, speech versus silence discrimination is first performed using spectral flatness, which is a measure of the energy distribution across frequencies. Spectral Flatness (SF) is computed frame-wise using the following equation, with the same frame length and hop length mentioned in the previous section.
\begin{equation}
\text{SF} = \frac{\exp \left( \frac{2}{N_{\text{FFT}}} \sum_{k=1}^{N_{\text{FFT}}/2} \ln |S_k| \right)}{\frac{2}{N_{\text{FFT}}} \sum_{k=1}^{N_{\text{FFT}}/2} |S_k|}
\end{equation}
where $N_{\text{FFT}}$ is the order of the Fourier transform and $|S_k|$ denotes the magnitude in the $k$th frequency bin of the speech signal.

The Spectral Flatness (SF) values in the voiced, unvoiced, and silence regions are analyzed. Experimental analysis show that silence regions tend to have a flatter spectrum compared to speech. Based on visual inspection threshold of 0.75 are set for the SF, classifying segments with an SF value of 0.75 or higher as silence regions. Based on the length of these silence regions, the break indices are assigned as follows:
\begin{equation}
	\text{Break Index} =
	\begin{cases} 
		1, & \text{if } l < 80 \text{ ms} \\
		2, & \text{if } 80 \text{ ms} \leq l < 290 \text{ ms } \\
		3, & \text{if } l \geq 290 \text{ ms }
	\end{cases}
\end{equation}

Here, Break Index values 1, 2, and 3 correspond to short, medium, and long inter-word
silences, respectively.

\subsection{Labeling Pitch Contours}
\label{sec:pc}
The pitch period and fundamental frequency (F0) calculation are essential for capturing prosodic characteristics in speech. While computationally cheaper algorithms for pitch period estimation exist, the group delay-based algorithm \cite{anushiyarachel15_interspeech} is employed due to the offline nature of our processing, where accuracy is prioritized over computational efficiency. The algorithm works by identifying the Glottal Closure Instants (GCIs), which mark the boundaries of individual glottal cycles. The difference between consecutive GCIs gives the pitch period \( T_0 \). The fundamental frequency \( F_0 \), which represents the rate of vocal fold vibrations, is then calculated as the inverse of the pitch period, \( F_0 = \frac{1}{T_0} \). 

The GCI algorithm is primarily used to identify pitch marks and for pitch tracking, which are crucial for various speech processing tasks. One such technique is the time-domain pitch synchronous overlap and add (TD-PSOLA), which modifies the prosody of speech and requires accurate pitch mark estimation. Additionally, the estimation of GCIs plays a significant role in speech dereverberation, glottal source modeling, speech enhancement, and speech synthesis. Accurate GCI detection ensures the effective manipulation of prosodic features, leading to more natural and intelligible speech synthesis and processing.

To ensure precise estimation of \( T_0 \) and \( F_0 \), speech signals are analyzed using short overlapping frames, typically 20 ms in size with a 10 ms hop length. GCIs are detected within each frame, and the pitch period is derived.

\begin{figure}[!ht]
	\centerline{\includegraphics[height=7cm, width=8cm]{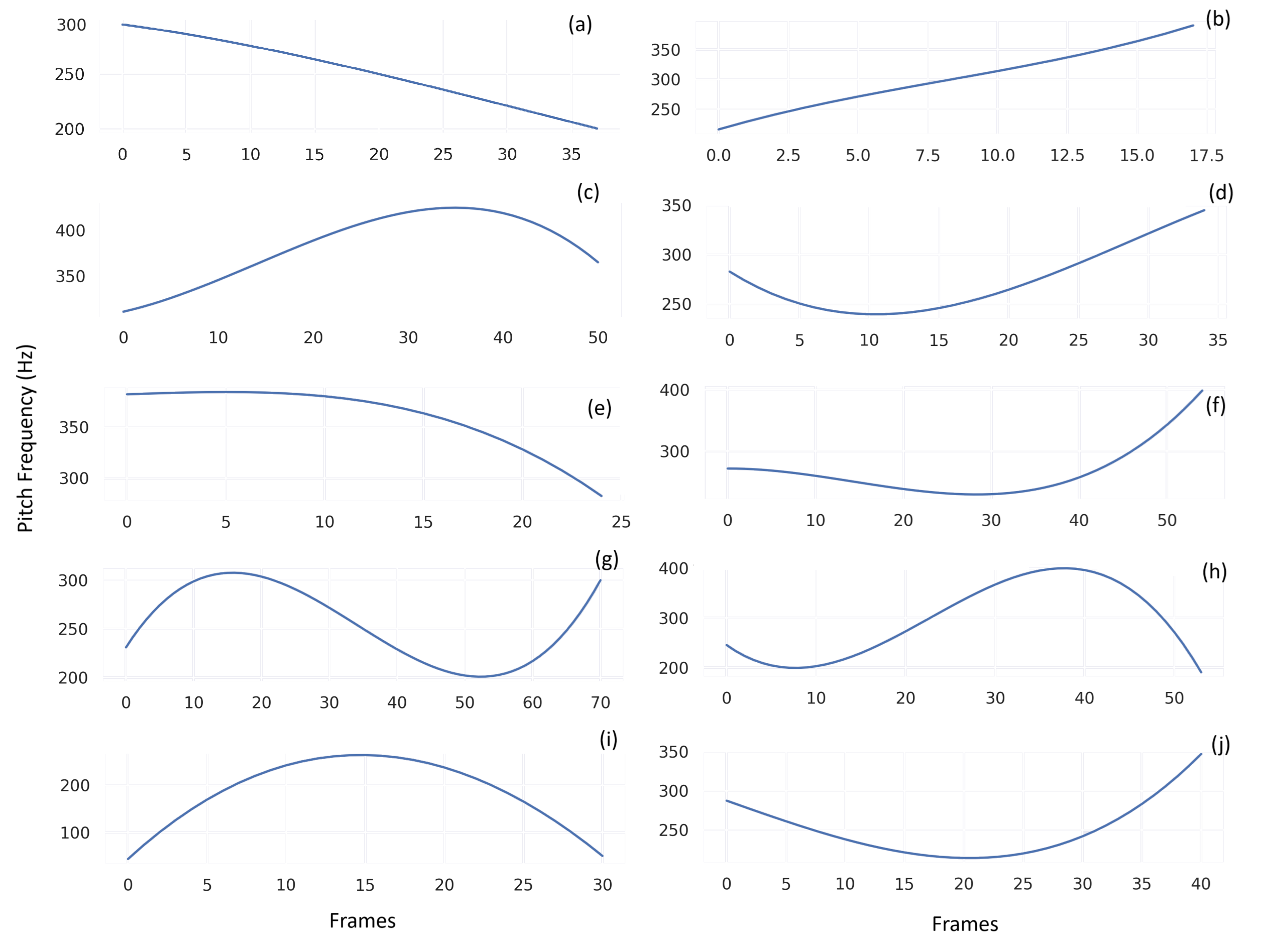}}
	\caption{Basic Pitch Contour Shapes Considered: (a)~L, (b)~H, (c)~HHL, (d)~LHH, (e)~HLL, (f)~LLH, (g)~HLH, (h)~LHL, (i)~Hat, (j)~Bucket}
	\label{fig:pc_label}
\end{figure}

Once the pitch contour, i.e., the time-varying \( F_0 \), is estimated, it undergoes syllable-level smoothing to reduce local fluctuations and abrupt changes. This is achieved using a polynomial of order 3, which fits a smooth curve to the pitch values over time, yielding a more natural representation of prosodic variations. Initially, eleven shapes of the pitch contour, namely L, H, HLL, HHL, LLH, LHH, HLH, LHL, Hat,Bucket, and flat, are identified, where L (low) represents a falling contour and H (high) represents a rising contour. To capture finer variations, each shape is further categorized into three classes based on their dynamic ranges, resulting in a total of 31 pitch contour classes. The dynamic ranges of the pitch frequency are represented with an S (small), M (medium), or B (big) prefix, corresponding to a range of 10--60 Hz, 60--100 Hz, and above 100 Hz, respectively. If the dynamic range is less than 10 Hz, the contour is classified as flat. The basic pitch contour shapes (except flat) are portrayed in figure \ref{fig:pc_label}.

\section{Performance Evaluation}
\label{sec:perf}
\subsection{Speech Corpus}
\label{sec:corpus}
The speech corpus used for analysis consists of read speech data from
both text-to-speech (TTS) and automatic speech recognition (ASR) resources for five
Indian languages: Indian English, Tamil, Hindi, Gujarati, and Malayalam. For each
language, a total of 6 hours of data is used, comprising 4 hours for training and 2
hours for evaluation.

The training data (4 hours per language) is constructed as follows:
(i) 1 hour of clean speech obtained from the IITM IndicTTS corpus \cite{iitm_indic_tts},
(ii) 1 hour of clean speech drawn from the IITM INX-SpeakerHub (multispeaker ASR) corpus
\cite{mary2024inx}, and
(iii) 2 hours of noise-augmented speech, created by adding car, babble, street, and
white noise (0.5 hours each) at 0, 5, and 10~dB SNR levels to the IndicTTS corpus. These noises are from the NOISEX-92\cite{varga1993noisex} and the NOIZEUS datasets\cite{hu2007speech}. The noise augmentation is
applied to assess the robustness of the proposed system under non-ideal acoustic
conditions.

The IndicTTS data is recorded by professional male and female
speakers in a controlled studio environment to maintain consistent voice quality and
pitch characteristics. The ASR data originates from the INX-SpeakerHub corpus, which contains speech from a large number of native Indian speakers
across multiple languages and provides substantial speaker diversity. All recordings
are mono-channel, sampled at 48~kHz, with 16-bit resolution, and stored in WAV format.
This corpus design enables systematic evaluation across multiple languages, speaker
conditions, and acoustic environments, while remaining suitable for low-resource
prosody annotation tasks.

\subsection{Accuracy of Segmentation}
\label{sec:seg_acc}
\begin{figure}[!htbp]
\includegraphics[width=0.45\linewidth]{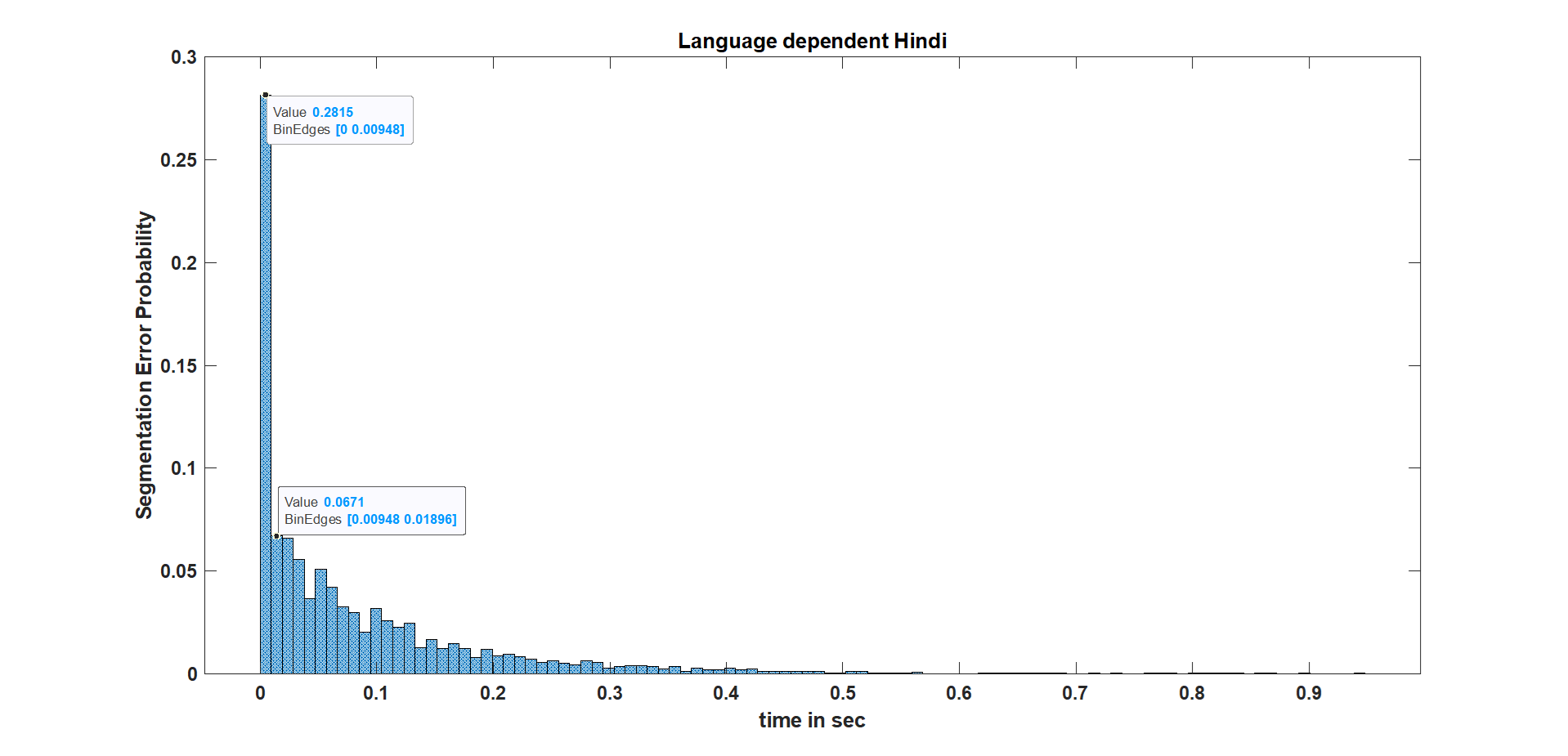}
\hfill
\includegraphics[width=0.45\linewidth]{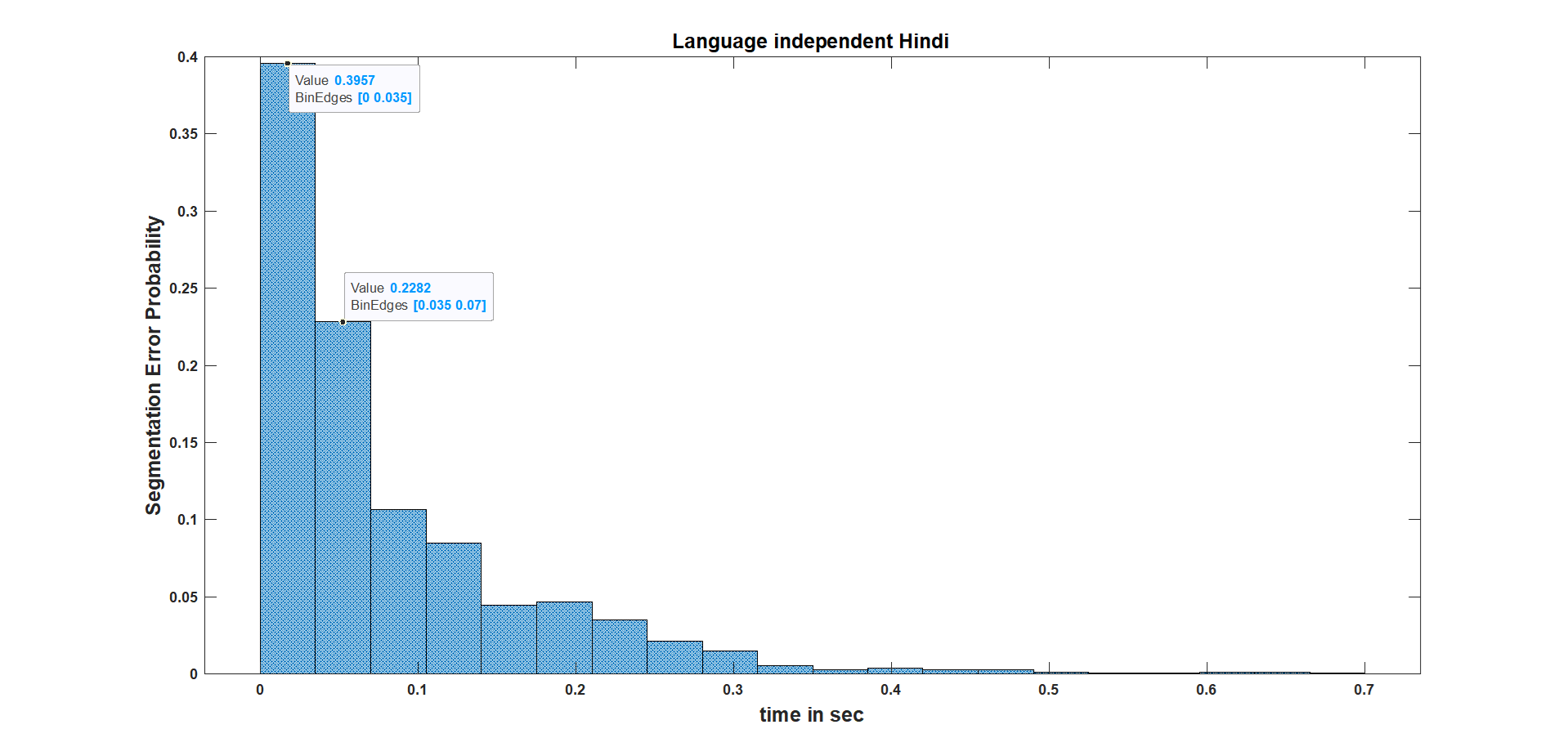}

\vspace{6pt}

\includegraphics[width=0.45\linewidth]{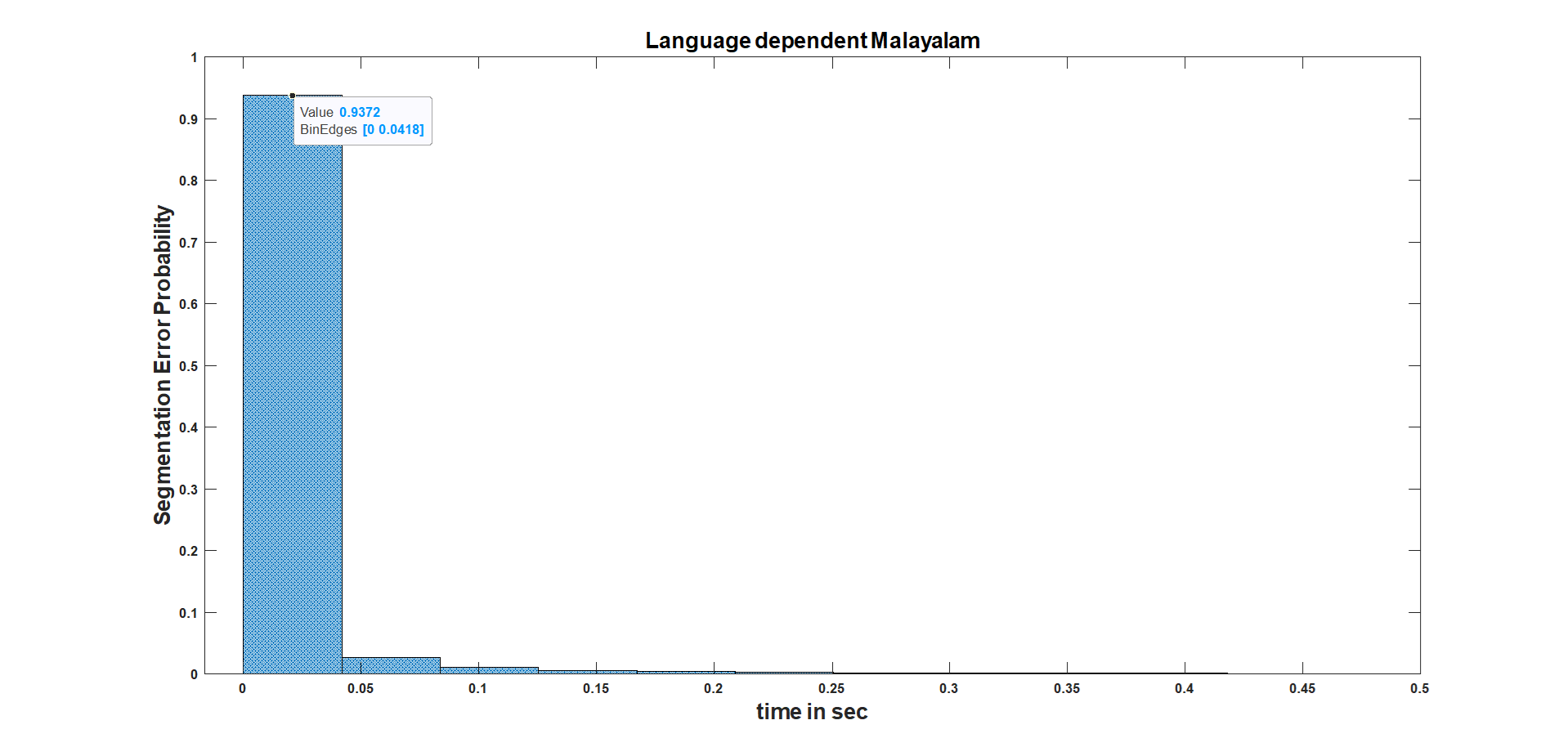}
\hfill
\includegraphics[width=0.45\linewidth]{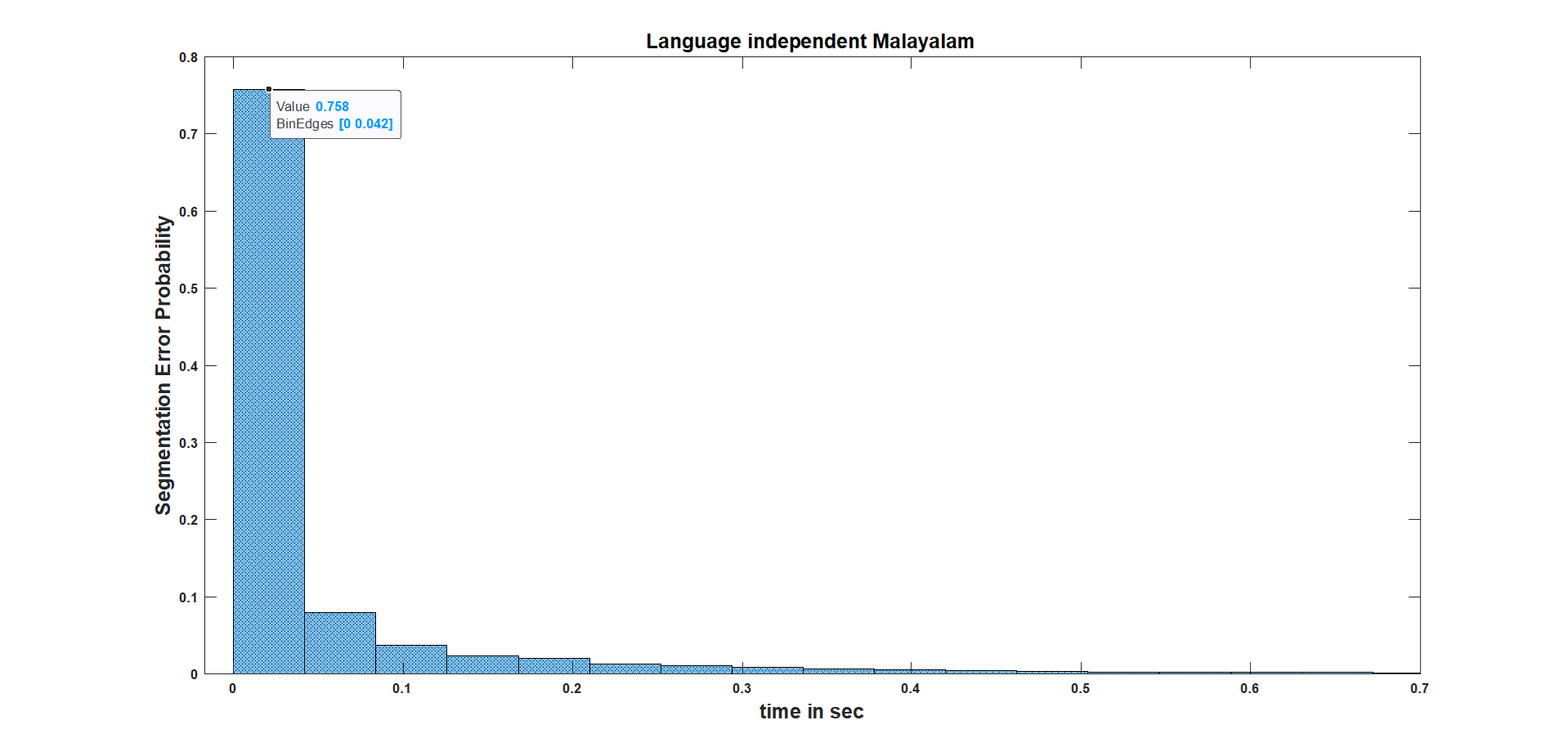}

\vspace{6pt}

\includegraphics[width=0.45\linewidth]{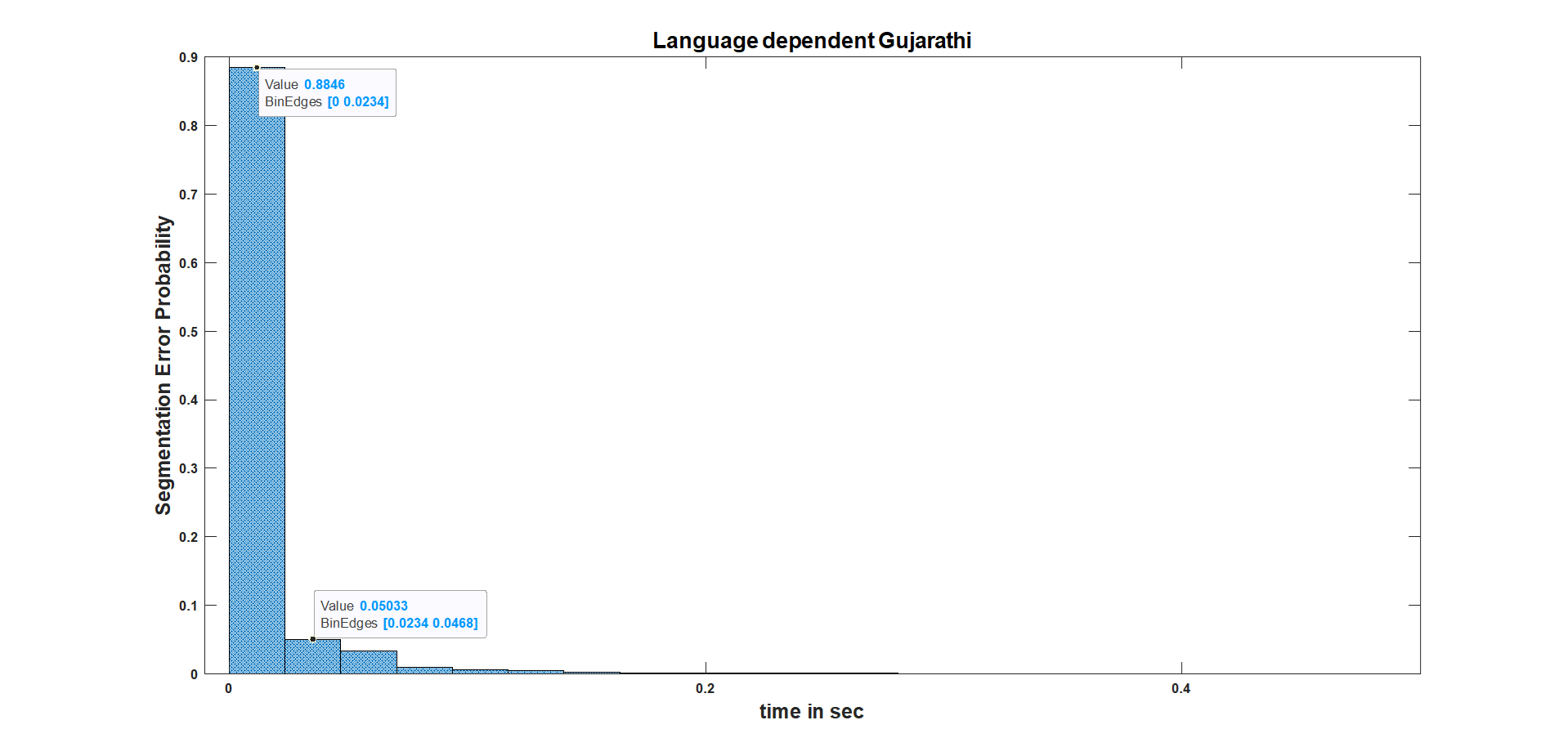}
\hfill
\includegraphics[width=0.45\linewidth]{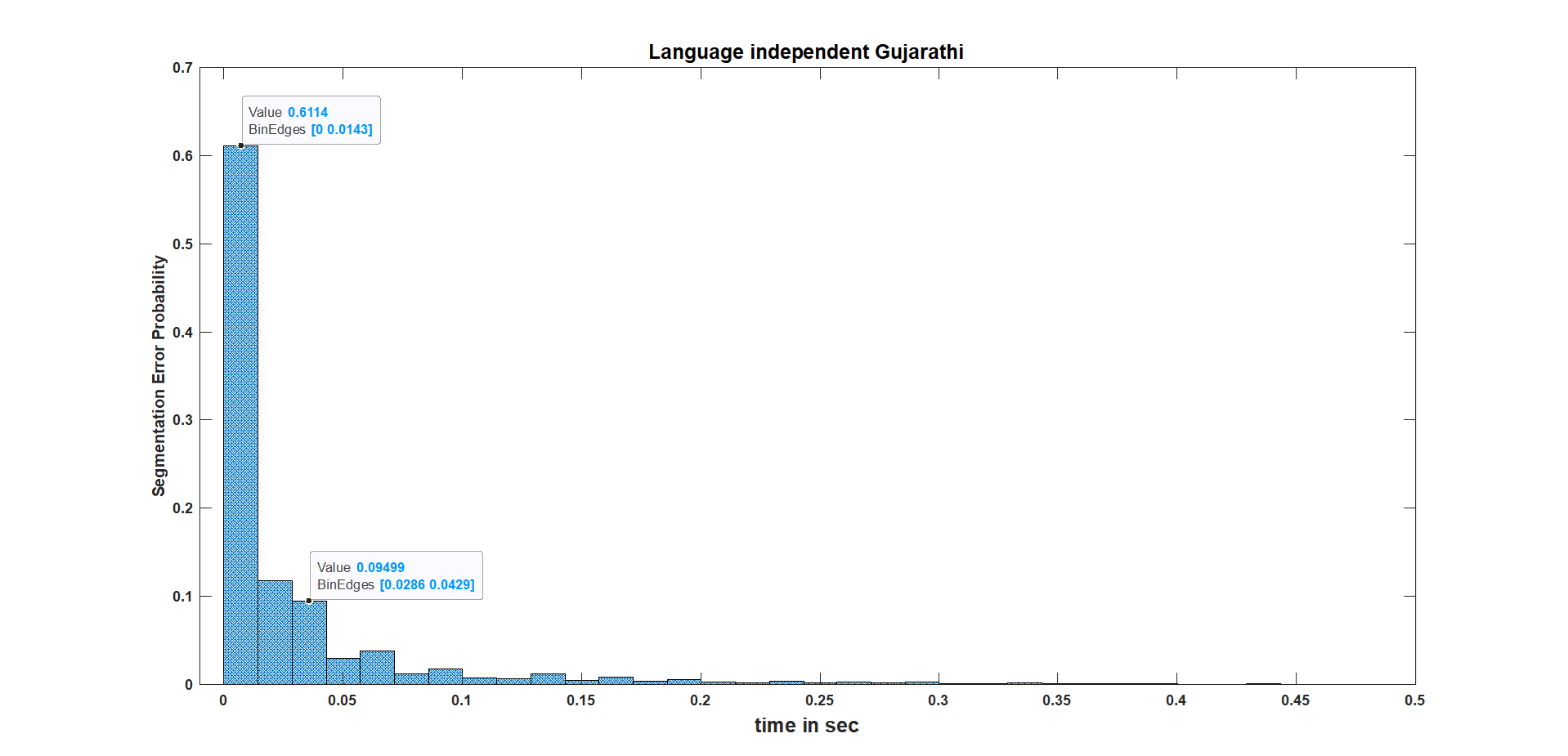}

\caption{Comparison of Segmentation Error for Language-Independent and Language-Dependent Models}
\label{fig:ld_li_10imgs}
\end{figure}

The accuracy with which the tool segments the given speech signals using the HMM-based forced-Viterbi alignment procedure is evaluated at the phoneme level, as syllable and word-level segmentations are derived from the phoneme-level segmentation. To evaluate accuracy, the segmentation error is calculated as the time difference between the manually derived duration for each phoneme and the corresponding duration obtained from the tool. The distribution of the segmentation errors for each language when language-dependent and language-independent models are used is shown in figure \ref{fig:ld_li_10imgs}.

It can be observed that ninety percent of segments in the language-dependent models across all five languages exhibit errors of less than 10ms. On average, the error lies between 10 to 30ms for the five languages when using both language-dependent and language-independent models. There is marginal difference of 10\% error observed in language-independent models compared to language-dependent models arises from the use of shared phonetic models across languages, which leads to some generalization errors. Nevertheless, this trade-off is deemed acceptable given the increased flexibility and efficiency offered by language-independent models, which are essential for accommodating multiple languages effectively.
\begin{figure}[!htbp]
\centering

\includegraphics[width=0.45\linewidth]{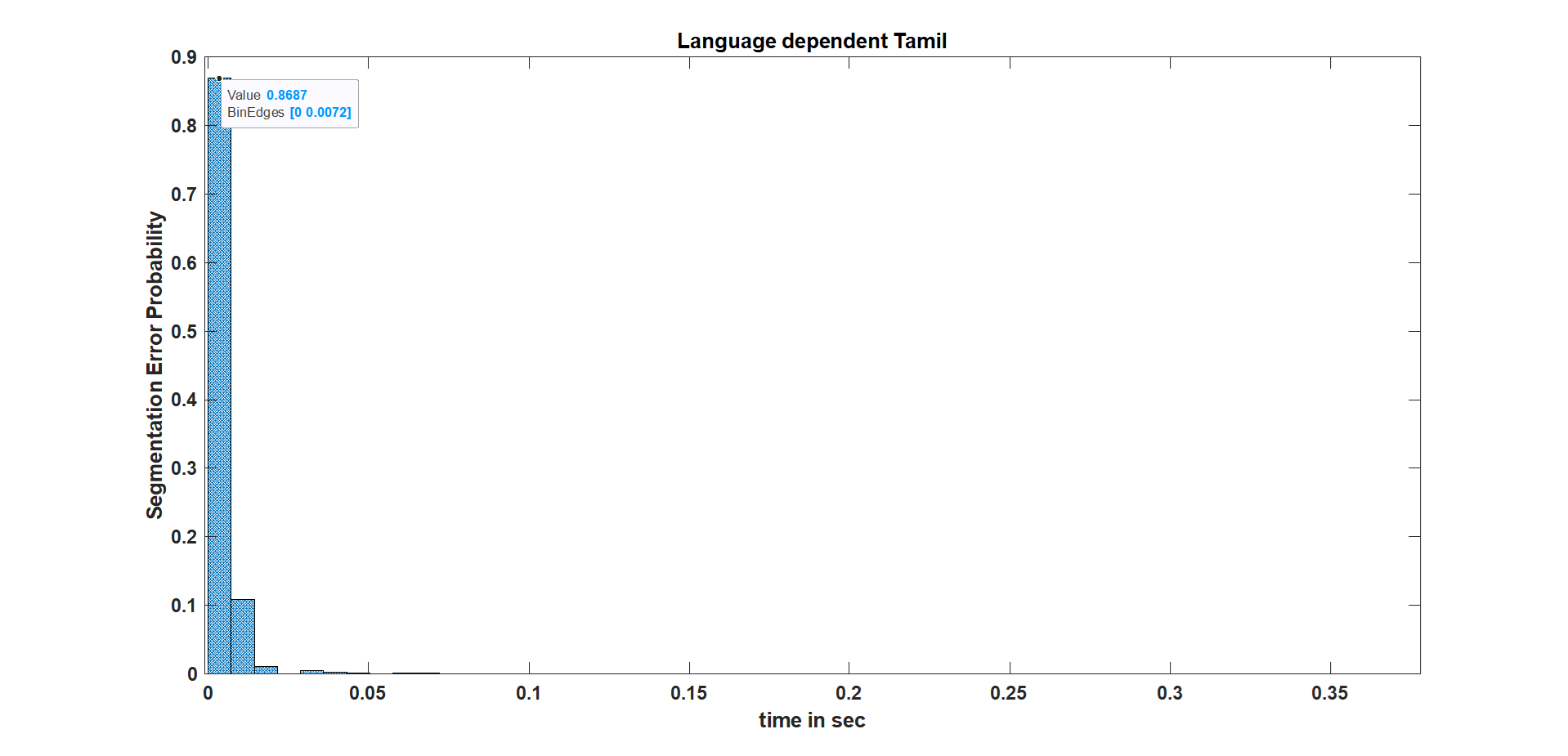}
\hfill
\includegraphics[width=0.45\linewidth]{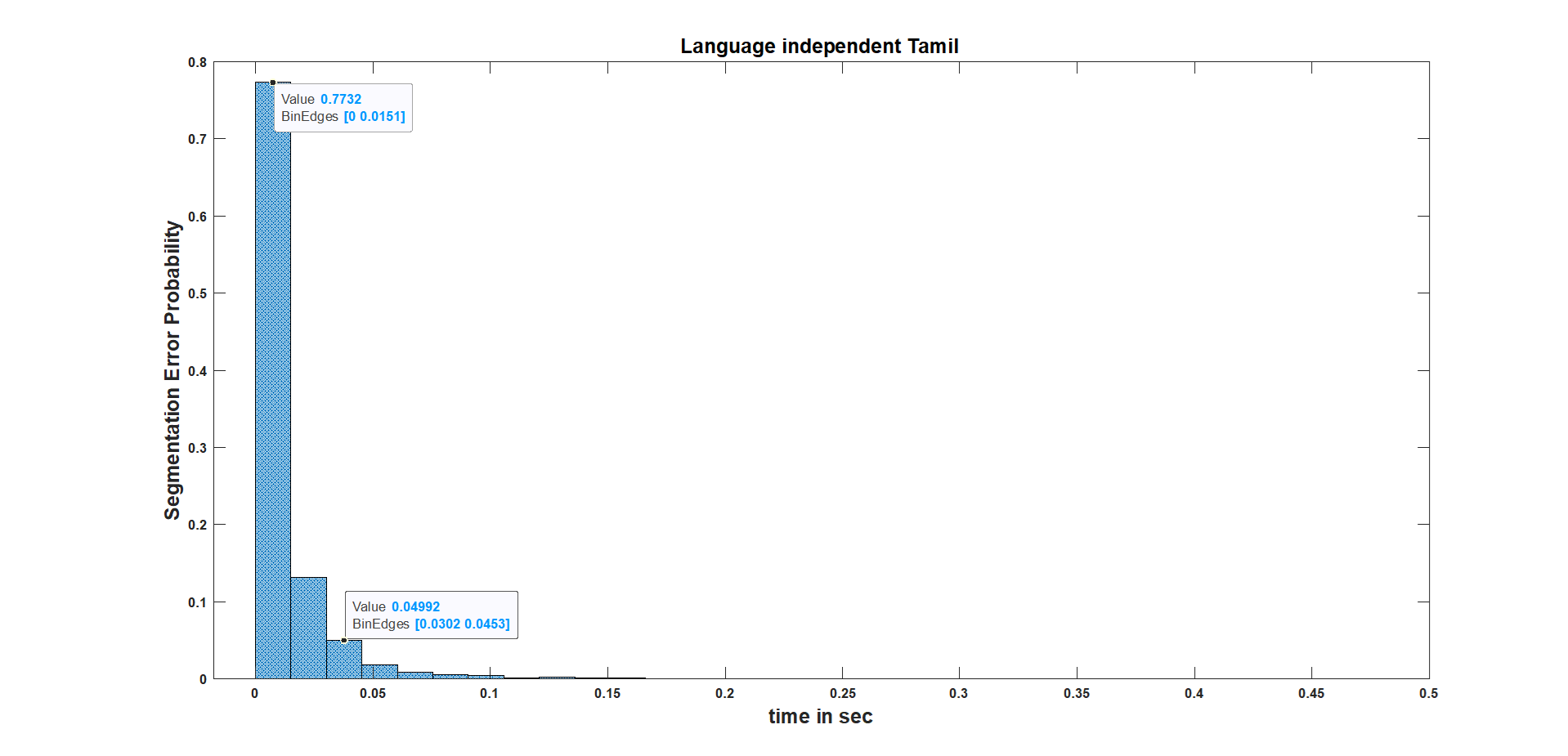}

\vspace{6pt}

\includegraphics[width=0.45\linewidth]{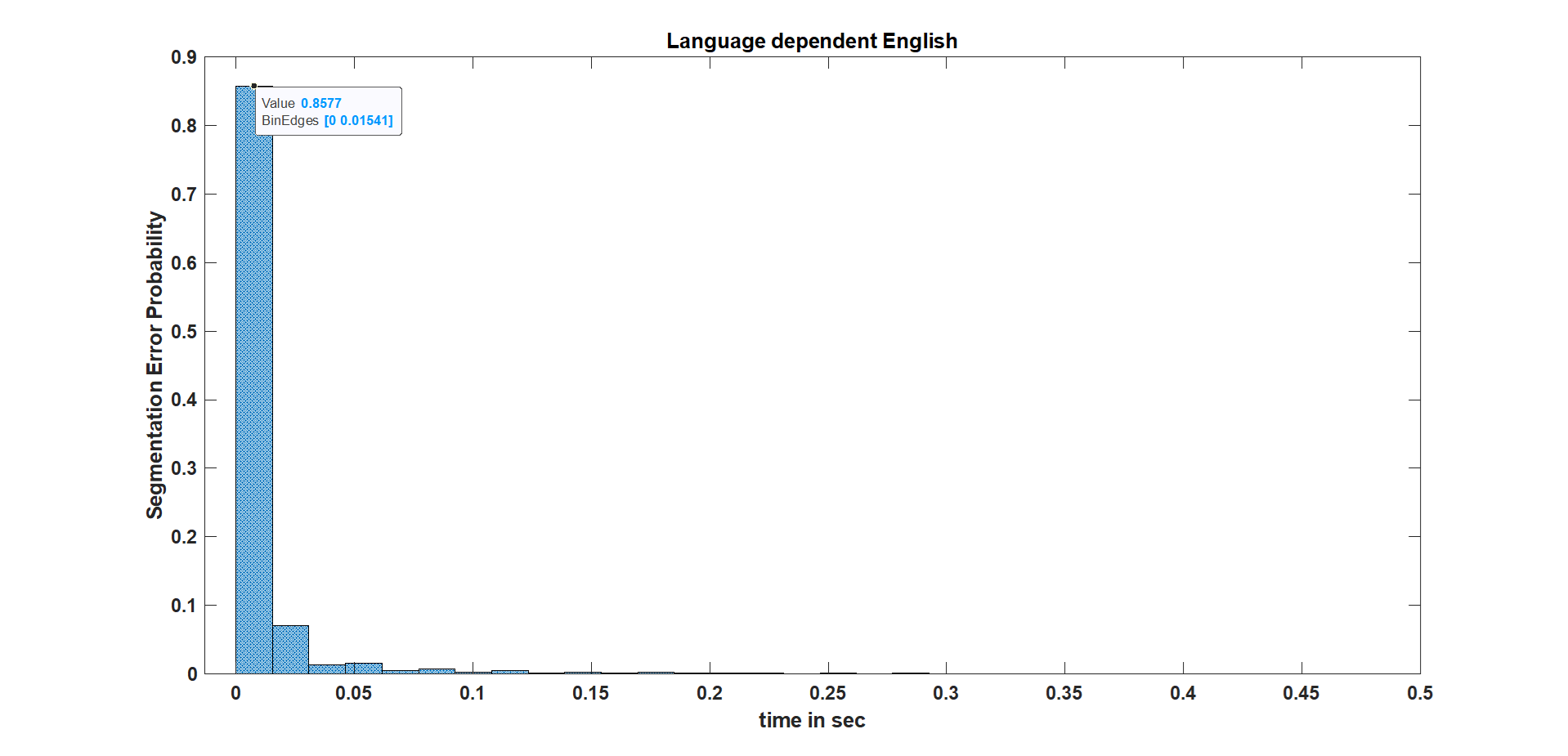}
\hfill
\includegraphics[width=0.45\linewidth]{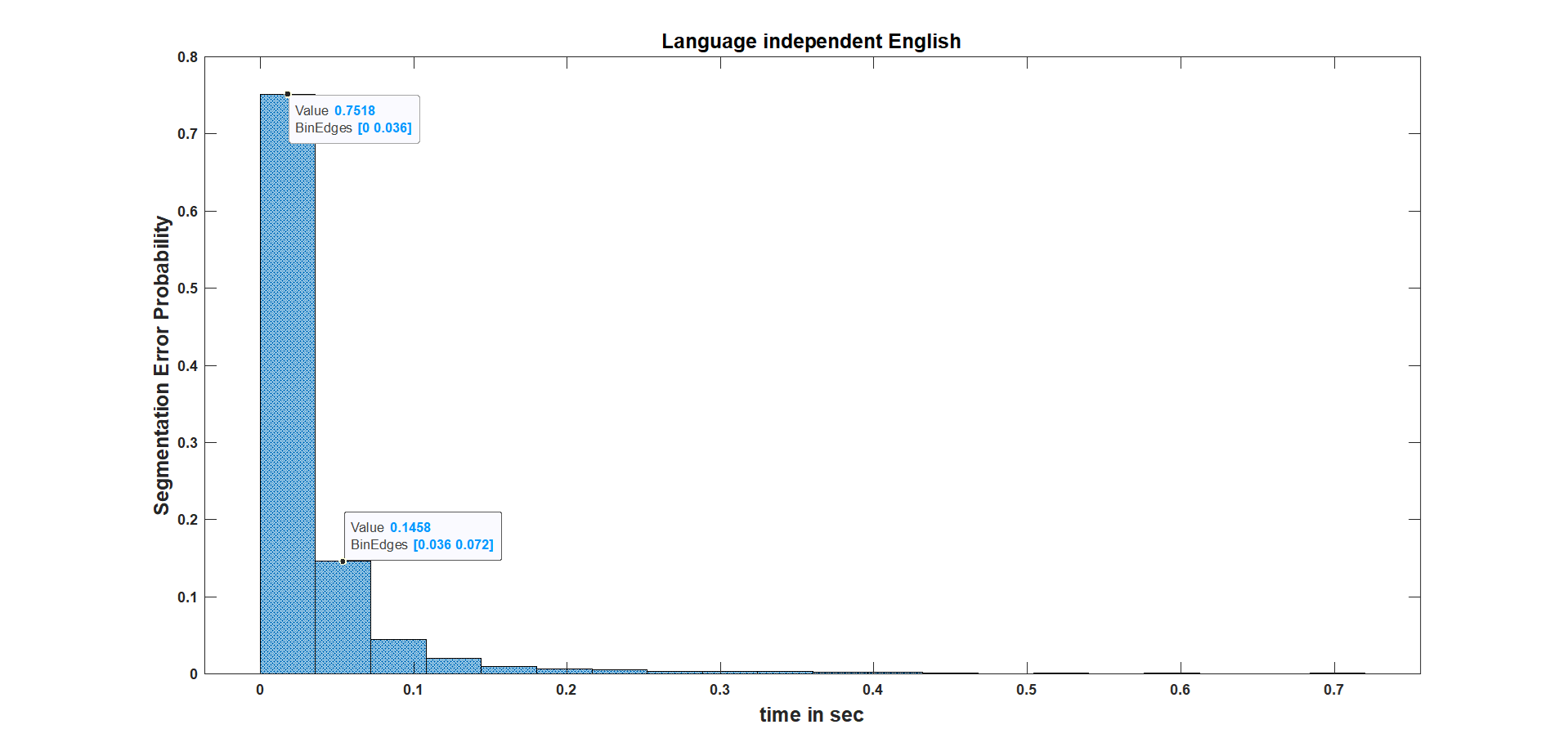}

\vspace{6pt}
\end{figure}







\newpage
\subsection{Accuracy of Break Indices}
\label{sec:bi_acc}
\begin{figure}[!htbp]
\centering

\begin{minipage}[c][5cm][c]{0.3\linewidth}
\centering
\includegraphics[width=\linewidth,height=5cm,keepaspectratio]{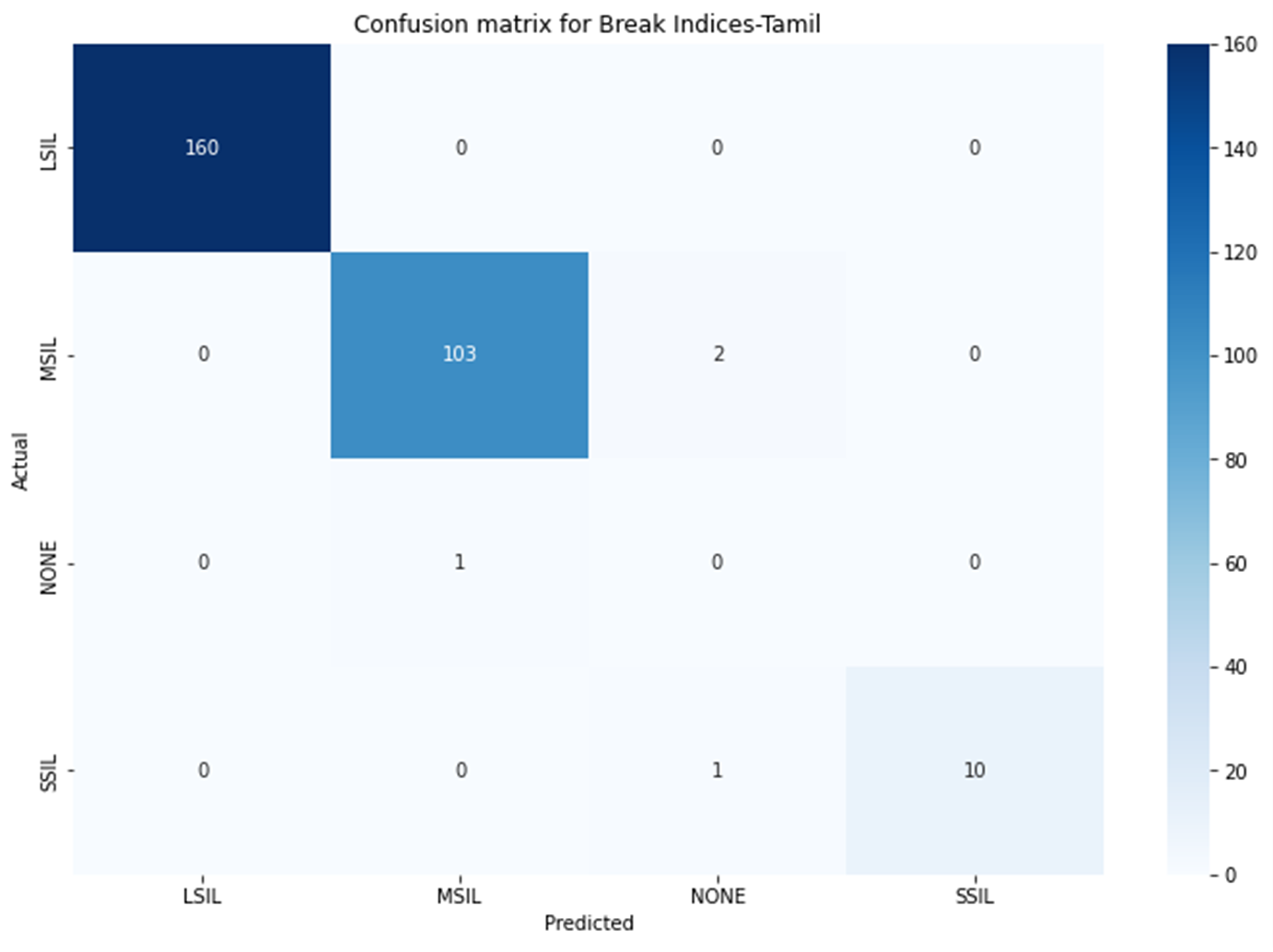}
\end{minipage}
\hfill
\begin{minipage}[c][5cm][c]{0.3\linewidth}
\centering
\includegraphics[width=\linewidth,height=5cm,keepaspectratio]{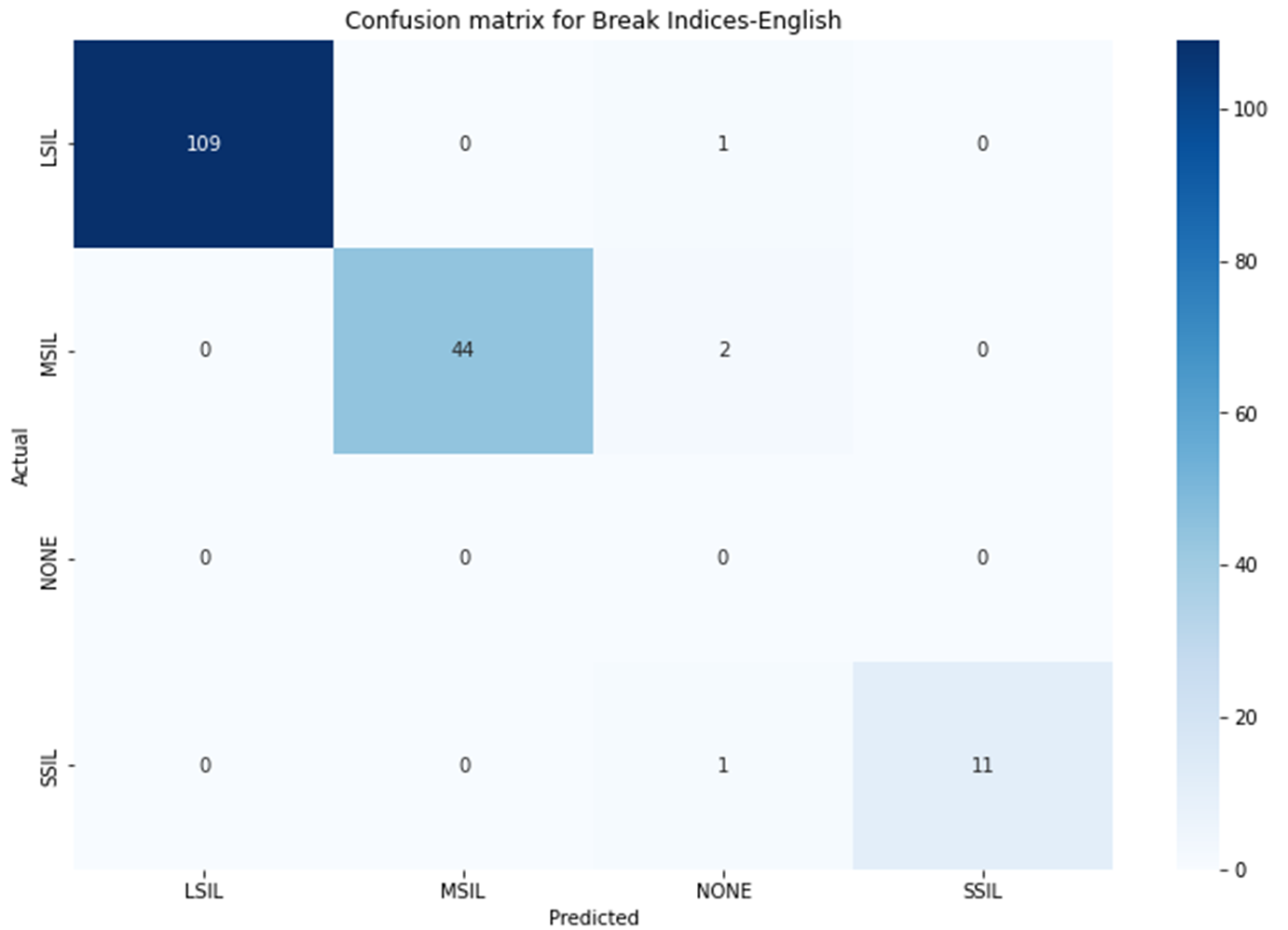}
\end{minipage}
\hfill
\begin{minipage}[c][5cm][c]{0.3\linewidth}
\centering
\includegraphics[width=\linewidth,height=5cm,keepaspectratio]{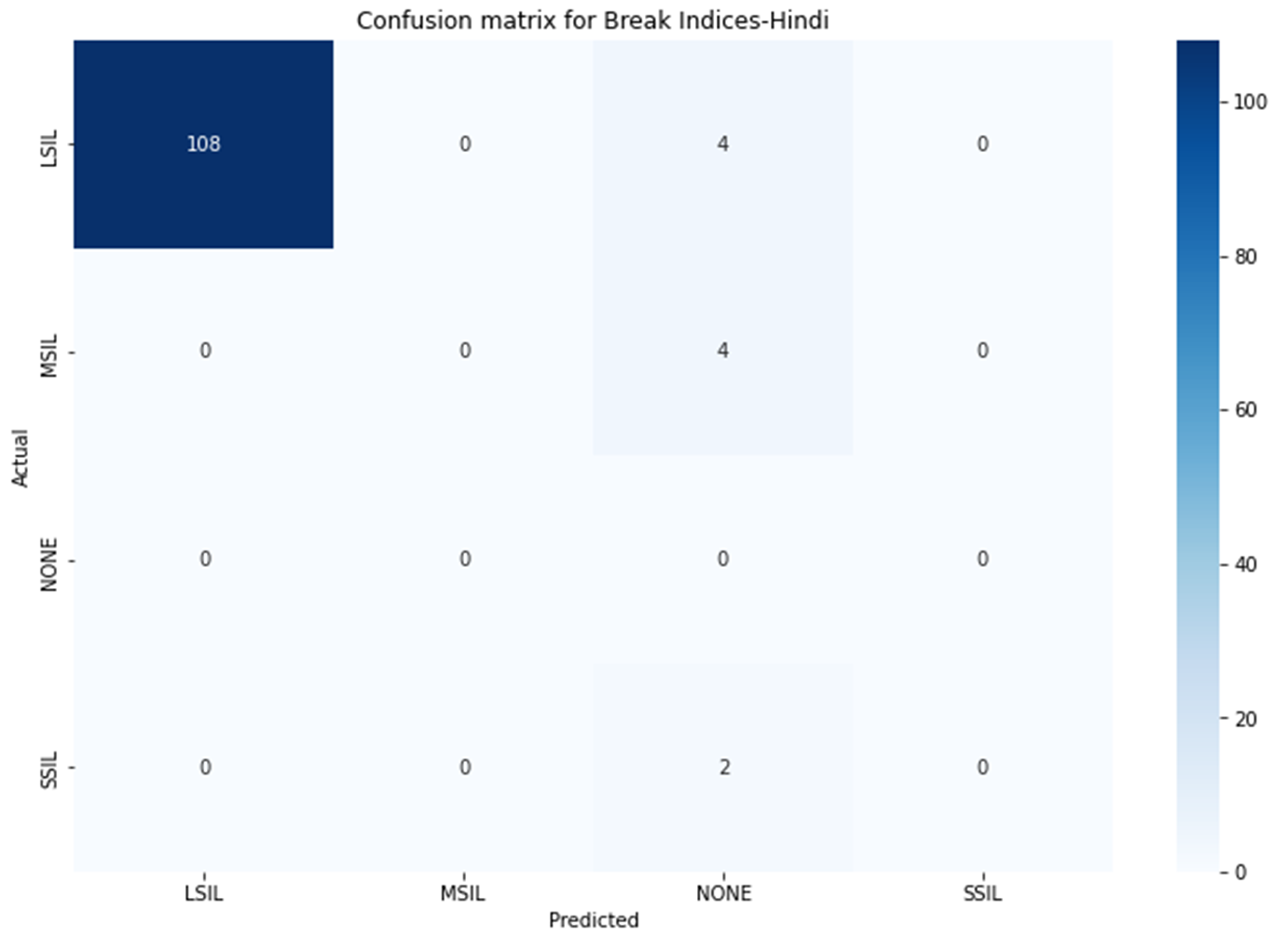}
\end{minipage}


\makebox[\linewidth][c]{%
\begin{minipage}[c][5cm][c]{0.3\linewidth}
\centering
\includegraphics[width=\linewidth,height=5cm,keepaspectratio]{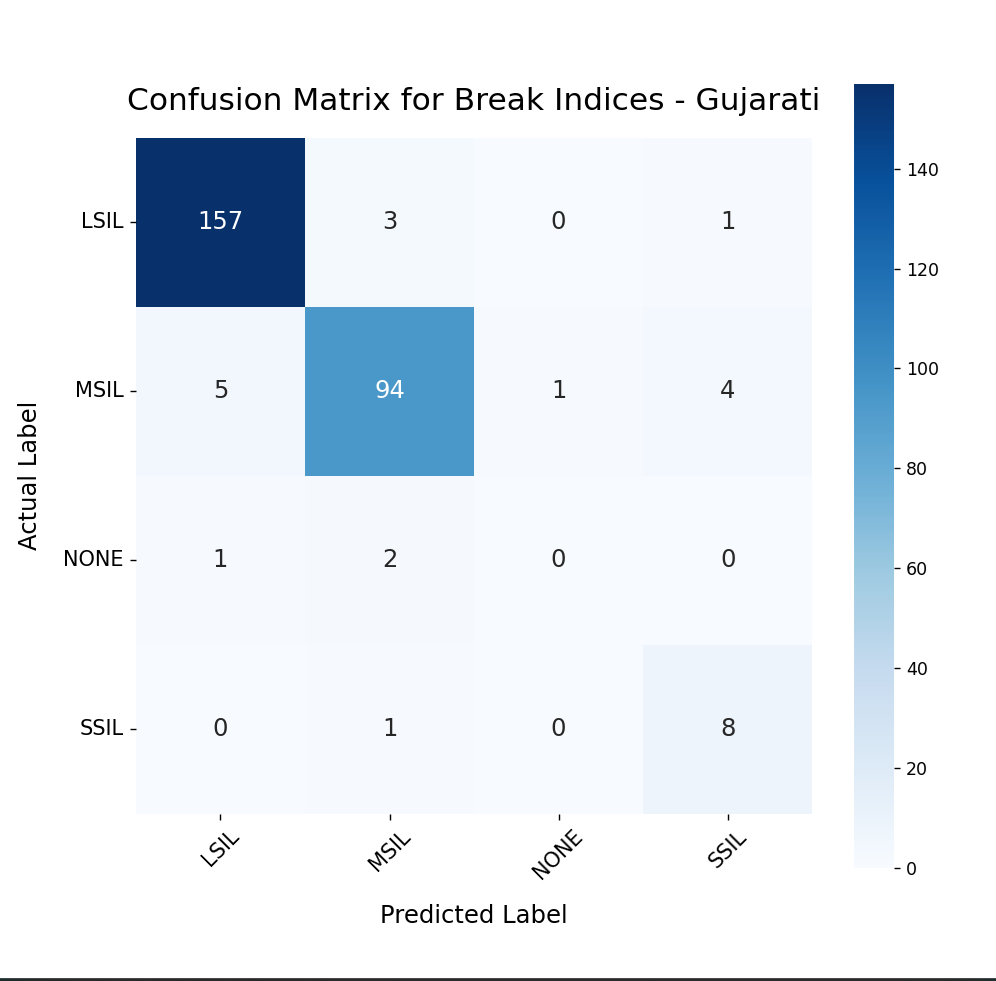}
\end{minipage}
\hspace{1cm}
\begin{minipage}[c][5cm][c]{0.3\linewidth}
\centering
\includegraphics[width=\linewidth,height=5cm,keepaspectratio]{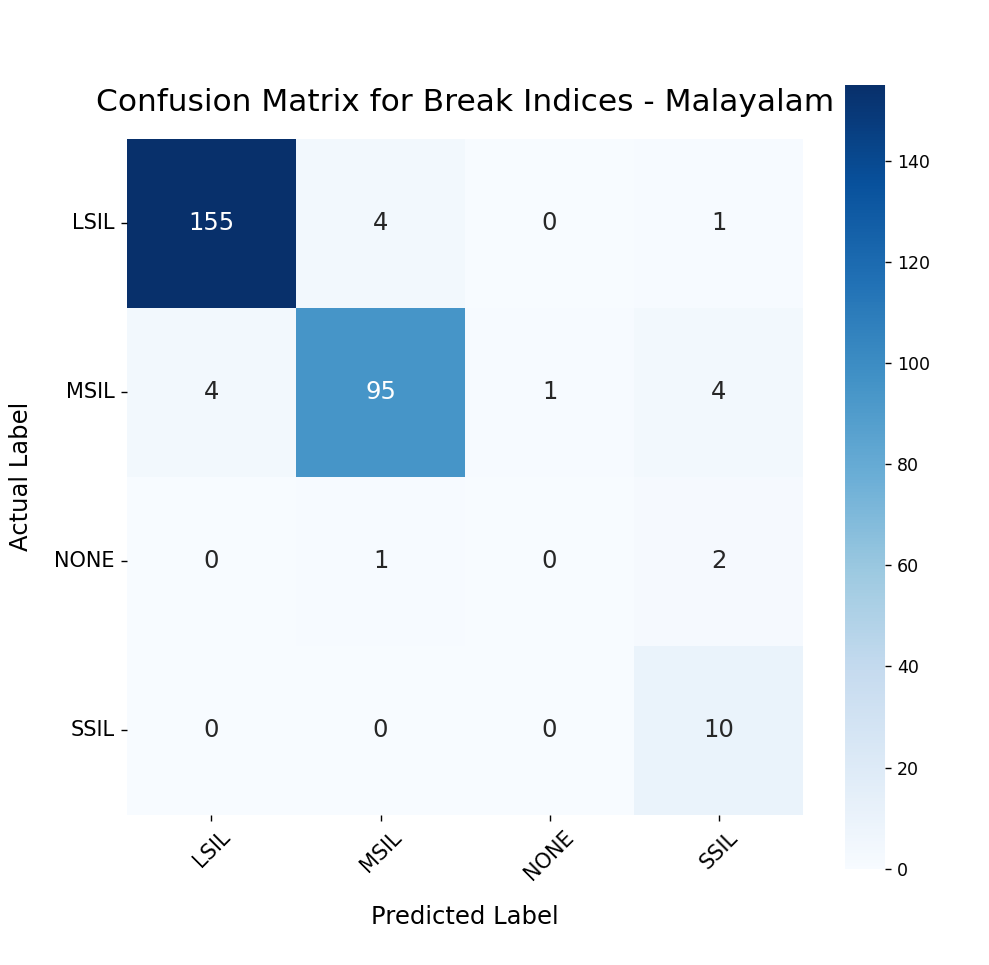}

\end{minipage}
}

\caption{Confusion matrices for the identification of Break Indices:
(a) Tamil, (b) Indian English, (c) Hindi, (d) Gujarati, and (e) Malayalam.}
\label{fig:bi_conf}
\end{figure}

To assess the accuracy of break indices derived from the tool, these were compared with the manual annotations. 
As shown in Table \ref{tab:prosody-results}, the break indices are derived with an accuracy of around 95\%. It is observed that in the remaining 5\% of the silence regions, the boundaries detected based on the spectral flatness is inaccurate resulting mostly in the incorrect identification of a break index of 2 as 3 or 1. Further, some of the silence regions occur before or between stops consonants and fricatives were not considered as silence. This is reflected in the confusion matrices in figure \ref{fig:bi_conf}. While Tamil, Indian English,
Gujarati, and Malayalam exhibit comparable performance, slightly lower accuracy is observed for Hindi  
attributed to the influence of fricatives. Silence preceding fricatives tends to have a lower Spectral Flatness (SF) value, which may contribute to the decreased accuracy in Hindi.

%

\subsection{Accuracy of Pitch Contour Labeling}
\label{sec:pc_acc}
As in the previous sections, the accuracy of the pitch contour labels is assessed by comparing the manual annotations with those derived from the tool. Manual labeling involves a careful analysis of the audio signals, where expert annotators listen to the speech and identify pitch contours by visually examining the waveform and their pitch frequency. The annotators mark significant points, such as rise and fall, to capture the contour patterns present in the speech. It is observed that the designed rules capture the shape of the pitch contours with an accuracy of approximately 99\%, as summarized in Table~\ref{tab:prosody-results}. The LLH label has a high error rate for English, while the L label has a high error rate for both Tamil and Hindi. Gujarati shows relatively higher confusion for the HL label, whereas Malayalam exhibits higher errors for the LH label.


\subsection{Performance Across Languages and Noise Conditions}
To analyze the impact of different training data sources, we consider three
individual training conditions: TTS-only, ASR-only, and noise-augmented-only training.
Models trained exclusively on clean TTS data achieve slightly higher
clean-speech performance, with language-dependent phoneme segmentation accuracies of
approximately 95--96\%, pitch contour correlations close to 0.99, and break index
F1-scores in the range of 0.96--0.97, but with reduced robustness under noisy conditions.
ASR-only training yields moderate and stable performance (Seg-LD $\approx$ 92--93\%,
BI-F1 $\approx$ 0.93--0.94), while noise-augmented-only training improves robustness at
the cost of a small reduction in clean prosodic accuracy (Seg-LD $\approx$ 90--92\%,
BI-F1 $\approx$ 0.91--0.93). The combined training strategy adopted in this work
provides a balanced trade-off between clean-speech accuracy and noise robustness,
resulting in the stable performance reported in Table~\ref{tab:prosody-results}. As expected, noise-augmented training results in a minor degradation in clean-speech
accuracy, offset by improved robustness under noisy conditions. The results reported in Table~\ref{tab:prosody-results} correspond to models trained
using a combined dataset consisting of 1~hour of clean TTS speech, 1~hour of
multispeaker ASR speech, and 2~hours of noise-augmented speech, as described in \ref{sec:corpus}.

\begin{table}[!ht]

\centering
\small

\begin{tabular}{lccccccc}
\multicolumn{8}{p{0.95\linewidth}}{\footnotesize\textit{Abbreviations:}
W/Utt = average words per utterance;
Seg-LD = language-dependent phoneme segmentation accuracy (\%);
Seg-LI = language-independent phoneme segmentation accuracy (\%);
Pitch-$r$ = pitch contour correlation;
BI-F1 = Break Index F1-score.} \\
\hline
\textbf{Lang.} &
\textbf{Hours} &
\textbf{\#Utts} &
\textbf{W/Utt} &
\textbf{Seg-LD} &
\textbf{Seg-LI} &
\textbf{Pitch-$r$} &
\textbf{BI-F1} \\
\hline
Eng & 4T/2E & 4200 & 6.4 & 94.6 & 85.1 & 0.98 & 0.95 \\
Tam & 4T/2E & 4100 & 5.9 & 95.2 & 86.0 & 0.98 & 0.96 \\
Hin & 4T/2E & 4050 & 6.7 & 91.8 & 82.4 & 0.97 & 0.91 \\
Guj & 4T/2E & 3950 & 6.1 & 93.7 & 84.6 & 0.98 & 0.96 \\
Mal & 4T/2E & 4000 & 5.6 & 94.1 & 85.0 & 0.98 & 0.93 \\
\hline
\end{tabular}
\caption{Prosody accuracy across languages with noise-augmented training data.}
\label{tab:prosody-results}
\end{table}

\section{Influence of pitch contour on Language Identification Using the AutoProsody Tool}
\label{sec:lid}
This study aims to check whether the frequency of pitch contours has any influence on language identification. Specifically, the current work attempts to identify the language of a given word based on syllable-level pitch contours, as prosodic information varies across languages. For this task, Tamil, Hindi, Indian English are considered. Initially, monosyllabic words from IITM IndicTTS corpus\cite{iitm_indic_tts} , comprising 1500 monosyllabic words from the  Tamil, Hindi, Indian English data. Pitch contour labels are estimated from 90\% of these words, and the normalized frequency of occurrence for each contour in each language is computed. For the remaining 10\% of the words, the previously computed normalized frequency of occurrence is assigned as a score based on the pitch contour.

This analysis is extended to bisyllabic, trisyllabic, quadsyllabic, and pentasyllabic words, where the normalized frequency of occurrence of a contour is computed for each syllable category (e.g., bisyllabic or trisyllabic). When identifying the language of a word, the cumulative score obtained across all syllables is considered, with the language having the highest cumulative score being identified as the language of the word.

\begin{figure}[htbp]
    \centerline{\includegraphics[height=5cm, width=8cm]{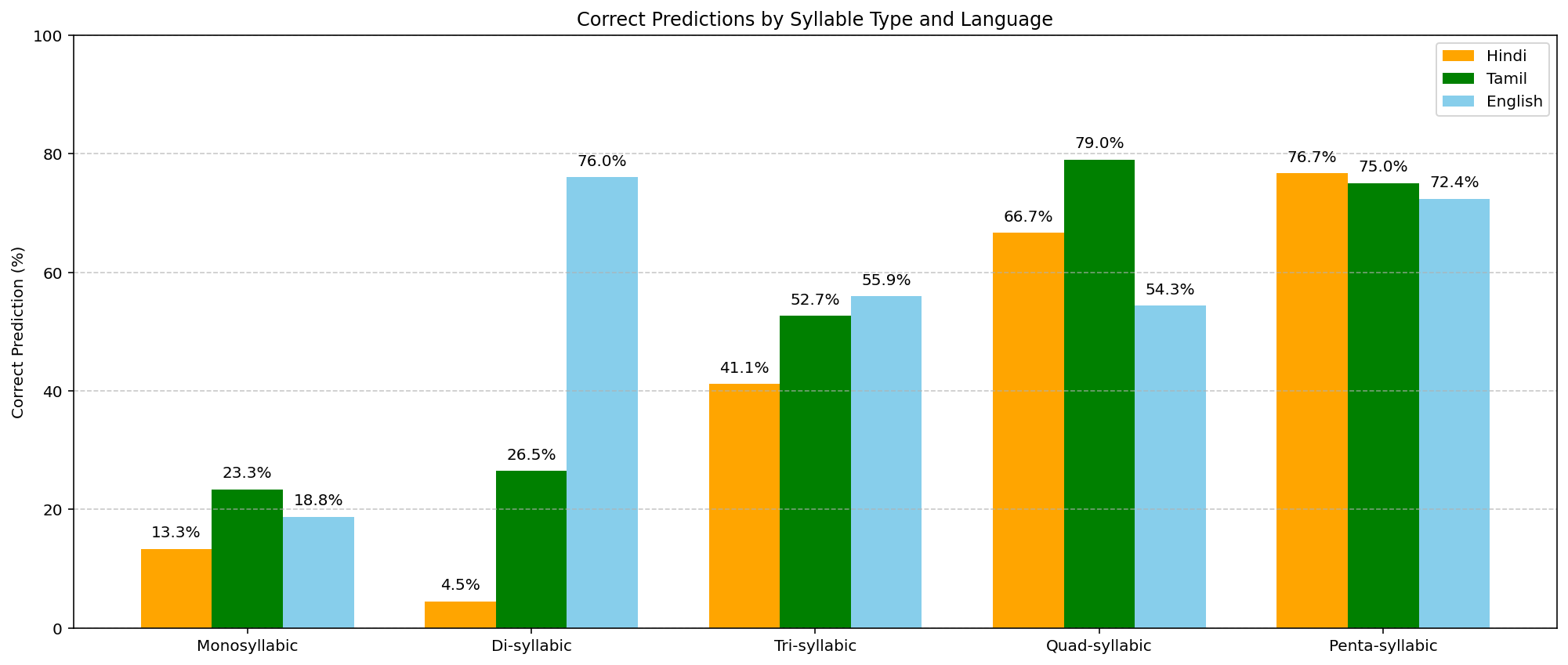}}
    \caption{Language Identification Accuracy for Mono, Bi, Tri, Quad, and Penta-syllabic words}
    \label{fig:lid}
\end{figure}

Figure~\ref{fig:lid} illustrates the precision of language identification in different syllable lengths for Hindi, Tamil, and Indian English. The results reveal that language identification accuracy generally improves with an increase in the syllable count. For Hindi, the model struggles with monosyllabic and disyllabic words, achieving only 13.3\% and 4.5\% accuracy, respectively. However, performance improves significantly for longer words, with accuracy reaching 41.1\% for trisyllabic, 66.7\% for quadsyllabic, and 76.7\% for pentasyllabic words. Tamil shows a more balanced trend, maintaining an accuracy greater than 23\% even for monosyllabic words and steadily increasing to 79. 0\% for quadsyllabic words and 75. 0\% for pentasyllabic words. English shows a distinct pattern, with high accuracy for disyllabic words at 76.0\%, moderate accuracy for tri- and pentasyllabic words (55.9\% and 72.4\% respectively), and relatively low performance for monosyllabic (18.8\%) and quadsyllabic (54.3\%) words. These patterns suggest that syllable length plays a crucial role in the model's ability to correctly identify the language, likely because of the increased linguistic cues available in longer words.

\section{Conclusion}
\label{sec:conc}
The proposed AutoProsody tool provides rich prosodic features for five languages, namely Indian English, Tamil, Hindi, Malayalam and Gujarati. The performance of each module of the tool has been compared with manual annotations, and the accuracy has been computed, demonstrating its effectiveness in capturing phonetic and prosodic details. While the current work focused on these five languages, the tool can be easily extended to other languages by retraining the phoneme HMMs and incorporating automatic speech recognition (ASR) systems for the new languages.
Additionally, the modular design of the tool allows for adaptability to various prosodic feature extraction tasks, making it a versatile tool for speech processing across different linguistic contexts. The general syllabification algorithm used in both language-dependent and language-independent models ensures that the tool can handle diverse language structures with minimal degradation in performance. Future work could explore applying this tool to low-resource languages, as well as refining the pitch contour labeling for better accuracy in tonal languages. The AutoProsody tool has the potential to significantly contribute to multilingual speech processing and analysis in a variety of applications, from language identification to speech synthesis and recognition.

\section*{Acknowledgments}
The current work is carried out as a part of the project titled, ``Prosody Modeling'', under the sub-project of the NLTM BHASHINI project, titled, ``Speech technologies in Indian languages'', funded by the Ministry of Electronics and Information Technology, Government of India, with reference number, 11(1)/2022-HCC(TDIL).

\bibliographystyle{plain}
\section*{Data Availability}
The data supporting the findings of this study are publicly available. The demo page of the AutoProsody tool can be accessed at: \url{https://speech.snuchennai.edu.in/demos.html}. The prosody-annotated datasets used in this work are available at: \url{https://github.com/speech-lab-snuchennai/Prosody-annotated-data}.

\section*{Funding}
The authors declare that no funds, grants, or other support were received during the preparation of this manuscript.

\section*{Corresponding Author}
Corresponding author: Preethi Thinakaran (preethit@snuchennai.edu.in)

\end{document}